\documentclass[pre,superscriptaddress,reprint,showpacs,onecolumn,notitlepage,nobalancelastpage]{revtex4-2}

\usepackage{graphicx}
\usepackage{tikz}
\usepackage{amsmath}
\usepackage{amssymb}
\usepackage{color}
\usepackage{float}
\usepackage{epstopdf}
\usepackage[normalem]{ulem}
\usepackage{xfrac}

\usepackage{pgfplots,pgfplotstable}
\usepackage{soul}
\usepackage{accents}
\usepackage{nameref}
\usepackage{balance}
\usepackage{upgreek}
\usepackage[all]{nowidow}
\usepackage{comment}
\usepackage{dsfont}
\usepackage{cleveref}
\usepackage{array} 				
\usepackage{booktabs} 			
\usepackage{ctable}
\newcolumntype{q}[1]{>{\setlength{\parindent}{0em}}p{#1}}

\crefformat{section}{\S#2#1#3}
\crefformat{subsection}{\S#2#1#3}
\crefformat{subsubsection}{\S#2#1#3}

\makeatletter
\newcommand*{\balancecolsandclearpage}{
  \close@column@grid
  \clearpage
  \twocolumngrid
}
\makeatother

\usetikzlibrary[shadings]
\usetikzlibrary{shapes}
\usetikzlibrary{snakes}
\usepackage{bm}
\usepgfplotslibrary{fillbetween}
\tikzset{
    axis break gap/.initial=10mm
}
\usetikzlibrary{patterns}

\pgfplotsset{compat=newest}

\sethlcolor{white}

\AtBeginDocument{
\heavyrulewidth=.08em
\lightrulewidth=.05em
\cmidrulewidth=.03em
\belowrulesep=.65ex
\belowbottomsep=0pt
\aboverulesep=.4ex
\abovetopsep=0pt
\cmidrulesep=\doublerulesep
\cmidrulekern=.5em
\defaultaddspace=.5em
}

\begin{document}

\title{Maximum Electro-Momentum Coupling in Piezoelectric Metamaterial Scatterers}

\author{Jeong-Ho Lee}
\author{Zhizhou Zhang}
\author{Grace X. Gu}
\email[]{ggu@berkeley.edu}
\affiliation{Department of Mechanical Engineering, University of California, Berkeley, CA 94720}

\begin{abstract}
Analogous to electromagnetic bianisotropy, engineered piezoelectric metamaterials can possess electro-momentum coupling between the macroscopic momentum and electric stimuli. This indicates the applicability of piezoelectric metamaterials for wave scattering with an extra design degree of freedom, in the same way as acoustic materials with Willis coupling between the macroscopic momentum and strain. To fully utilize this novel bianisotropy, we derive for the first time tight theoretical bounds on the effect of electro-momentum coupling on scatterers based on energy conservation and combining two acoustic and electromagnetic polarizability tensors to analyze passive bianisotropic scatterers under both acoustic and electromagnetic waves. Our derived bounds are verified by comparing them with analytical scattering solutions. Results show that the bianisotropic scattering performance can be of the same order as the non-bianisotropic terms via the aid of electro-momentum coupling, even for simple designs with small Willis coupling. We further reveal the possibility of utilizing electro-momentum coupling for tunable scattering-cloaking applications of the subwavelength-sized scatterers. This sheds light on the promising potential of piezoelectric metamaterials for tunable scattering devices in which bianisotropy can be modulated by external electric stimuli.
\end{abstract}

\maketitle

The emergence and application of metamaterials have created new opportunities for manipulating dynamic material responses, such as electromagnetic waves through artificial magnetism with negative-index, subwavelength focusing, and invisibility cloaking {\cite{smith2000negative,pendry2006controlling,leonhardt2009broadband,schurig2006metamaterial,pendry2000negative}}. Metamaterials with engineered spatial distribution have been found to possess Willis coupling in elastodynamics, when their property fluctuations are averaged out with material homogenization. Analogous to the bianisotropy in electromagnetics {\cite{marques2002role,zhou2005saturation,asadchy2018bianisotropic,zheludev2012metamaterials,zhao2014recent,yu2014flat,glybovski2016metasurfaces}}, Willis coupling shows an additional coupling effect at the subwavelength scale between strain and macroscopic (homogenized) linear momentum, which is coupled only to velocity by mass density at the microscale {\cite{willis1981variational,milton2007modifications,norris2012analytical,muhlestein2016reciprocity}}. This direct analogy has motivated applications of these elastodynamic, bianisotropic metamaterials to manipulate elastic waves such as acoustic waves for inaudible cloaks, collimators, and superlenses {\cite{li2004double,fang2006ultrasonic,ambati2007surface,torrent2007acoustic,zhang2009focusing,zhang2011broadband,PhysRevB.92.104105,cummer2016controlling,quan2018maximum}}, in which Willis coupling represents the interaction between particle velocity and acoustic pressure. While this interaction has been, so far, considered as a higher-order perturbation {\cite{muhlestein2016reciprocity,sieck2017origins,koo2016acoustic}}, it is revealed that for acoustic scattering, the effect of Willis coupling can have the same order as the non-bianisotropic terms {\cite{quan2018maximum}}. However, previous work is limited only to metamaterials deformable by mechanical forces.

For piezoelectric metamaterials responding to electric fields, it has recently been discovered that engineered spatial distribution induces an additional bianisotropic phenomenon, called electro-momentum (EM) coupling, between elastodynamics and electrostatics. With material homogenization, this bianisotropy defines a novel cross-coupling between the macroscopic (homogenized) linear momentum and the electric field at the subwavelength scale {\cite{pernas2020symmetry,pernas2021electromomentum,muhafra2022homogenization,zhang2022rational}}. This EM coupling can be represented in the effective (homogenized) constitutive equations for internal variables of metamaterials as 
\begin{equation}
\label{constitutiveeq_macro}
\begin{pmatrix}
\langle\bm{\sigma}\rangle \\ \langle\bm{D}\rangle \\ \langle\bm{\mu}\rangle
\end{pmatrix}
=
\begin{bmatrix}
\widetilde{\bm{C}} & \widetilde{\bm{B}}^{T} & \widetilde{\bm{S}} \\
\widetilde{\bm{B}} & -\widetilde{\bm{A}} & \widetilde{\bm{W}} \\
\widetilde{\bm{S}}^{\dagger} & \widetilde{\bm{W}}^{\dagger} & \widetilde{\bm{\rho}}
\end{bmatrix}
\begin{pmatrix}
\langle\bm{\varepsilon}\rangle \\ \langle\bm{\nabla}\phi\rangle \\ \langle\dot{\bm{u}}\rangle
\end{pmatrix}
\end{equation}
with the angle bracket $\langle\cdot\rangle$ denoting homogenization by ensemble averaging where $\bm{\sigma}$, $\bm{\mu}$, $\bm{\varepsilon}$, and $\bm{u}$ are stress, linear momentum, elastic strain, and displacement, respectively, with the over-dot denoting the time derivative. $\bm{D}$ and $\phi$ are electric displacement and electric potential, respectively, such that the electric field $\bm{E}=-\bm{\nabla}\phi$ from the quasi-electrostatic characteristic of piezoelectric materials. The over-tilde denotes effective properties where $\bm{C}$, $\bm{B}$, $\bm{A}$, and $\bm{\rho}$ are elastic, piezoelectric, dielectric, and density tensors, respectively. Note that the effective mass density is tensorized while the mass density is scalar, $\rho$, at the microscale. Here, $\widetilde{\bm{S}}$ and $\widetilde{\bm{W}}$ denote Willis and EM coupling, respectively, with adjoint operator $(\cdot)^{\dagger}$. One unprecedented feature of EM coupling is that the piezoelectricity allows control of this coupling effect through modulation of external electric stimuli $\bm{\mathcal{T}}_{e}$ as $\widetilde{\bm{W}}=\widetilde{\bm{W}}(\bm{\mathcal{T}}_{e})$ (i.e., post-fabrication modification of functionality), in contrast to typical metamaterials whose bianisotropy is fixed at the time of fabrication. This implies not only the applicability of piezoelectric metamaterials for manipulating elastic waves such as acoustic waves but also the tunability of the wave manipulation performance by external stimuli that can be considered as an extra design degree of freedom.

To fully explore the programmable bianisotropy of piezoelectric metamaterials, in this paper we derive tight theoretical bounds of EM coupling on wave scatterers based on energy conservation. Building on the fact that piezoelectric materials relate two fundamental fields (mechanical and electric), we utilize and combine two polarizability tensors to analyze passive bianisotropic scatterers under acoustic and electromagnetic waves. While the acoustic polarizability tensor relates acoustic pressure and velocity to acoustic monopole and dipole moments, the electromagnetic polarizability tensor links electric and magnetic fields to electric and magnetic dipole moments. Similar to Willis coupling, these bounds are determined solely by time frequency, and independent of the non-bianisotropic terms {\cite{quan2018maximum}}. Since both Willis and EM couplings are associated with linear momentum, those are not independent and interact with each other. Accordingly, our derived bounds show that along with the acoustic pressure-velocity coupling, the acoustic pressure-electromagnetic fields coupling can become of the same order as the non-bianisotropic terms as Willis and EM couplings assist one another to enhance the bianisotropy. Furthermore, our results show the possibility of using EM coupling for tunable cloaking.

Inducing a change in the direction and motion of waves due to a collision with another material object, scattering has been successfully used for manipulating waves, as obstacles or medium fluctuations of small dimensions can modify wave propagation in the medium. For acoustic waves, it has been studied that small metamaterial inclusions with the elastodynamic bianisotropy, Willis coupling, can have comparable scattering performance to non-bianisotropic scattering {\cite{quan2018maximum}}. This implies the applicability of EM coupling for general wave scatterers, not only for acoustic waves but also electromagnetic waves, with programmable performance. However, contrary to Willis coupling, the theoretical bounds of this coupling on general wave scattering have not been reported in previous literature. Thus, the performance space of piezoelectric metamaterial possessing EM coupling is poorly understood.

For subwavelength particles excited by an acoustic wave in a fluid, the scattering of this can be described by using the superposition of acoustic monopole $\mathcal{M}_{a}$ and dipole $\bm{\mathcal{D}}_{a}$ moments that are proportional to local acoustic pressure $p_{\text{loc}}$ and velocity $\bm{v}_{\text{loc}}$ for linear non-bianisotropic inclusions, respectively. These local fields are created by external excitations. In terms of Willis coupling {\cite{sieck2017origins,quan2018maximum}}, a general relation has been proposed using an acoustic polarizability tensor $\bm{\alpha}_{a}$ as 
\begin{equation}
\label{polarizability_a}
\begin{pmatrix}
\mathcal{M}_{a} \\ \bm{\mathcal{D}}_{a}
\end{pmatrix}
=
\bm{\alpha}_{a}
\begin{pmatrix}p_{\text{loc}} \\ \bm{v}_{\text{loc}}\end{pmatrix}
=
\begin{bmatrix} \alpha^{pp} & \bm{\alpha}^{pv} \\ \bm{\alpha}^{vp} & \bm{\alpha}^{vv}\end{bmatrix}
\begin{pmatrix}p_{\text{loc}} \\ \bm{v}_{\text{loc}}\end{pmatrix}
\end{equation}
in which the off-diagonal terms represent the effect of Willis coupling. Similarly, the electromagnetic response of bianisotropic metamaterials subjected to an electromagnetic wave is described by electric $\bm{\mathcal{D}}_{e}$ and magnetic $\bm{\mathcal{D}}_{m}$ dipole moments that are related to local electric $\bm{E}_{\text{loc}}$ and magnetic $\bm{H}_{\text{loc}}$ fields, respectively, such that a general relation can be represented via an electromagnetic polarizability tensor $\bm{\alpha}_{em}$ as
\begin{equation}
\label{polarizability_em}
\begin{pmatrix}
\bm{\mathcal{D}}_{e} \\ \bm{\mathcal{D}}_{m}
\end{pmatrix}
=
\bm{\alpha}_{em}
\begin{pmatrix}\bm{E}_{\text{loc}} \\ \bm{H}_{\text{loc}}\end{pmatrix}
=
\begin{bmatrix} \bm{\alpha}^{EE} & \bm{\alpha}^{EH} \\ \bm{\alpha}^{HE} & \bm{\alpha}^{HH}\end{bmatrix}
\begin{pmatrix}\bm{E}_{\text{loc}} \\ \bm{H}_{\text{loc}}\end{pmatrix}.
\end{equation}

Based on these general bianisotropic relations, energy conservation provides a fundamental constraint on the acoustic and electromagnetic polarizability tensors. For the sake of generality as well as simplicity, we proceed in a two-dimensional manner such that there is no variation in the out-of-plane direction from the 2D plane. The multipole expansion is applied using 2D Green's function for the acoustic scalar potential and the electromagnetic vector potential.

In the absence of energy sources, the passivity regarding energy conservation requires that net energy flux into a material region, $\partial\Omega$, must be equal to or greater than zero, which indicates that the scattered power must be equal to or less than the extinction power, with complex conjugates $(\cdot)^{*}$, as
\begin{equation}
\int{p_{s}\bm{v}_{s}^{*}}~d\partial\Omega \le -\int{(p_{s}\bm{v}_{in}^{*}+p_{in}\bm{v}_{s}^{*})}~d\partial\Omega
\end{equation}
for the acoustic part where $p_{in}$ and $p_{s}$ are the incident and scattered acoustic pressure, respectively, and $\bm{v}_{in}$ and $\bm{v}_{s}$ are the incident and scattered velocity, respectively, plus
\begin{equation}
\int{\bm{E}_{s}\times\bm{H}_{s}^{*}}~d\partial\Omega \le -\int{(\bm{E}_{s}\times\bm{H}_{in}^{*}+\bm{E}_{in}^{*}\times\bm{H}_{s})}~d\partial\Omega
\end{equation}
for the electromagnetic part where $\bm{E}_{in}$ and $\bm{E}_{s}$ are the incident and scattered electric field, respectively, and $\bm{H}_{in}$ and $\bm{H}_{s}$ are the incident and scattered magnetic field, respectively. As the scattered field is dominated by the monopole and dipole moments for subwavelength scatterers, these energy conservation inequalities can be rewritten as
\begin{subequations}
\label{passivity}
\begin{equation}
\big\vert\sqrt{2}\omega_{a}^{2}\mathcal{M}_{a}\big\vert^{2}+\big\vert ik_{a}\omega_{a}^{2}\bm{\mathcal{D}}_{a}\big\vert^{2} \le 8 \text{Im}\left[(\rho c_{a} \bm{v}_{in_0}^{*})\cdot(ik_{a}\omega_{a}^{2}\bm{\mathcal{D}}_{a})-p_{in_0}^{*}(\omega_{a}^{2}\mathcal{M}_{a})\right]
\end{equation}
\begin{equation}
\begin{aligned}
    & \Big\vert\frac{\omega_{em}^{2}}{c_{em}}(\bm{\mathcal{D}}_{e}^{xy}+\sqrt{2}\bm{\mathcal{D}}_{e}^{z})\Big\vert^{2}+\Big\vert \frac{\omega_{em}^{2}}{c_{em}^{2}}(\bm{\mathcal{D}}_{m}^{xy}+\sqrt{2}\bm{\mathcal{D}}_{m}^{z})\Big\vert^{2} \\
    &~~~~~~~~~~ \le 8 \text{Im}\left[\left(\frac{\omega_{em}^{2}}{c_{em}}(\bm{\mathcal{D}}_{e}^{xy}+\sqrt{2}\bm{\mathcal{D}}_{e}^{z})\right)\cdot\left(\frac{1}{\mathcal{Z}}(\bm{E}_{in_0}^{xy^*}+\frac{1}{\sqrt{2}}\bm{E}_{in_0}^{z^*})\right)+\left(\frac{\omega_{em}^{2}}{c_{em}^{2}}(\bm{\mathcal{D}}_{m}^{xy}+\sqrt{2}\bm{\mathcal{D}}_{m}^{z})\right)\cdot\left(\bm{H}_{in_0}^{xy^*}+\frac{1}{\sqrt{2}}\bm{H}_{in_0}^{z^*}\right)\right]
\end{aligned}
\end{equation}
\end{subequations}
with the subscript $0$ denoting quantities at the center of scatterers and the superscripts $xy$ and $z$ denoting the xy-plane and z-axis components of vectors, respectively. $i$ is the imaginary number, and $c_{a}$ and $c_{em}$ are the acoustic and electromagnetic velocities, respectively, and $k_{a}$ is the acoustic wavenumber, and $\mathcal{Z} = \sqrt{\mu_{em}/\epsilon_{em}}$ is the electromagnetic impedance with electric permittivity $\epsilon_{em}$ and magnetic permeability $\mu_{em}$. In this study, time-harmonic motions are assumed using the time convention $e^{-it\omega_{a}}$ and $e^{-it\omega_{em}}$ for the acoustic and electromagnetic fields, respectively, with time $t$, acoustic time frequency $\omega_{a}$, and electromagnetic time frequency $\omega_{em}$. Note that here all properties are of the background medium.

The usage of piezoelectric materials creates an interaction between elasticity and electrostatics. Hence, using the incident waves for the local quantities associated with external excitations, the polarizability tensors of (\ref{polarizability_a}) and (\ref{polarizability_em}) can be combined for piezoelectric metamaterials as
\begin{equation}
\label{polarizability_general}
\begin{pmatrix}
-\sqrt{2}\omega_{a}^{2}\mathcal{M}_{a} \\ ik_{a}\omega_{a}^{2}\bm{\mathcal{D}}_{a} \\ \omega_{em}^{2}\breve{\bm{\mathcal{D}}}_{e}/c_{em} \\ \omega_{em}^{2}\breve{\bm{\mathcal{D}}}_{m}/c_{em}^{2}
\end{pmatrix}
=
\bm{\alpha}'
\begin{pmatrix}
p_{in_0}/\sqrt{2} \\
\rho c_{a}\bm{v}_{in_0} \\
\breve{\bm{E}}_{in_0}/\mathcal{Z} \\
\breve{\bm{H}}_{in_0}
\end{pmatrix}
\end{equation}
where $\breve{\bm{\mathcal{D}}}_{e}=\bm{\mathcal{D}}_{e}^{xy}+\sqrt{2}\bm{\mathcal{D}}_{e}^{z}$, and $\breve{\bm{\mathcal{D}}}_{m}=\bm{\mathcal{D}}_{m}^{xy}+\sqrt{2}\bm{\mathcal{D}}_{m}^{z}$, and $\breve{\bm{E}}_{in_0}=\bm{E}_{in_0}^{xy}+\frac{1}{\sqrt{2}}\bm{E}_{in_0}^{z}$, and $\breve{\bm{H}}_{in_0}=\bm{H}_{in_0}^{xy}+\frac{1}{\sqrt{2}}\bm{H}_{in_0}^{z}$, with
\begin{equation}
\label{polarizability_normalized}
\bm{\alpha}' = 
\begin{bmatrix}
\alpha'^{pp} & \bm{\alpha}'^{pv} &  \bm{\alpha}'^{pE} & \bm{\alpha}'^{pH} \\
\bm{\alpha}'^{vp} & \bm{\alpha}'^{vv} & \bm{\alpha}'^{vE} & \bm{\alpha}'^{vH} \\
\bm{\alpha}'^{Ep} & \bm{\alpha}'^{Ev} & \bm{\alpha}'^{EE} & \bm{\alpha}'^{EH} \\
\bm{\alpha}'^{Hp} & \bm{\alpha}'^{Hv} & \bm{\alpha}'^{HE} & \bm{\alpha}'^{HH}
\end{bmatrix}
=
\begin{bmatrix}
-2\omega_{a}^{2}\alpha^{pp} & \frac{-\sqrt{2}\omega_{a}^{2}}{\rho c_{a}}\bm{\alpha}^{pv} & -\sqrt{2}\omega_{a}^{2}\mathcal{Z}\bm{\alpha}^{pE} & -\sqrt{2}\omega_{a}^{2}\bm{\alpha}^{pH} \\
i\sqrt{2}k_{a}\omega_{a}^{2}\bm{\alpha}^{vp} & \frac{ik_{a}\omega_{a}^{2}}{\rho c_{a}}\bm{\alpha}^{vv} & i k_{a}\omega_{a}^{2}\mathcal{Z}\bm{\alpha}^{vE} & ik_{a}\omega_{a}^{2}\bm{\alpha}^{vH} \\
\frac{\sqrt{2}\omega_{em}^{2}}{c_{em}}\bm{\alpha}^{Ep} & \frac{\omega_{em}^{2}}{\rho c_{a}c_{em}}\bm{\alpha}^{Ev} & \frac{\omega_{em}^{2}\mathcal{Z}}{c_{em}}\bm{\alpha}^{EE} & \frac{\omega_{em}^{2}}{c_{em}}\bm{\alpha}^{EH} \\
\frac{\sqrt{2}\omega_{em}^{2}}{c_{em}^{2}}\bm{\alpha}^{Hp} & \frac{\omega_{em}^{2}}{\rho c_{a}c_{em}^{2}}\bm{\alpha}^{Hv} & \frac{\omega_{em}^{2}\mathcal{Z}}{c_{em}^{2}}\bm{\alpha}^{HE} & \frac{\omega_{em}^{2}}{c_{em}^{2}}\bm{\alpha}^{HH}
\end{bmatrix}
\end{equation}
where $\bm{\alpha}'$ means a combined, normalized polarizability tensor. Then, summing (\ref{passivity}a-b) together and substituting the general relation of (\ref{polarizability_general}) gives a polarizability condition over bianisotropic particles as
\begin{equation}
\label{diag}
\text{Diag}\left[\frac{1}{4}(\bm{\alpha}'^{T^*}\bm{\alpha}')\right] \le \text{Diag}\left[i(\bm{\alpha}'^{T^*}-\bm{\alpha}')\right].
\end{equation}

In general bianisotropic cases with 2D-plane acoustic waves and electromagnetic waves of transverse electric (TE) polarization, the constraint (\ref{diag}) from energy conservation requires
\begin{subequations}
\label{maximum}
\begin{equation}
\label{maximum1}
\big\vert\alpha'^{vp}_{x}\big\vert^{2} + \big\vert\alpha'^{vp}_{y}\big\vert^{2} + \big\vert\alpha'^{Ep}_{x}\big\vert^{2} + \big\vert\alpha'^{Ep}_{y}\big\vert^{2} + \big\vert\alpha'^{Hp}_{z}\big\vert^{2} \le \frac{-8\text{Im}(1/\alpha'^{pp})-1}{\big\vert1/\alpha'^{pp}\big\vert^{2}} \le  16
\end{equation}

\begin{equation}
\label{maximum2}
\big\vert\alpha'^{pv}_{x}\big\vert^{2} + \big\vert\alpha'^{vv}_{yx}\big\vert^{2} + \big\vert\alpha'^{Ev}_{xx}\big\vert^{2} + \big\vert\alpha'^{Ev}_{yx}\big\vert^{2} + \big\vert\alpha'^{Hv}_{zx}\big\vert^{2} \le \frac{-8\text{Im}(1/\alpha'^{vv}_{xx})-1}{\big\vert1/\alpha'^{vv}_{xx}\big\vert^{2}} \le  16
\end{equation}

\begin{equation}
\label{maximum3}
\big\vert\alpha'^{pv}_{y}\big\vert^{2} + \big\vert\alpha'^{vv}_{xy}\big\vert^{2} + \big\vert\alpha'^{Ev}_{xy}\big\vert^{2} + \big\vert\alpha'^{Ev}_{yy}\big\vert^{2} + \big\vert\alpha'^{Hv}_{zy}\big\vert^{2} \le \frac{-8\text{Im}(1/\alpha'^{vv}_{yy})-1}{\big\vert1/\alpha'^{vv}_{yy}\big\vert^{2}} \le  16
\end{equation}

\begin{equation}
\label{maximum4}
\big\vert\alpha'^{pE}_{x}\big\vert^{2} + \big\vert\alpha'^{vE}_{xx}\big\vert^{2} + \big\vert\alpha'^{vE}_{yx}\big\vert^{2} + \big\vert\alpha'^{EE}_{yx}\big\vert^{2} + \big\vert\alpha'^{HE}_{zx}\big\vert^{2} \le \frac{-8\text{Im}(1/\alpha'^{EE}_{xx})-1}{\big\vert1/\alpha'^{EE}_{xx}\big\vert^{2}} \le  16
\end{equation}

\begin{equation}
\label{maximum5}
\big\vert\alpha'^{pE}_{y}\big\vert^{2} + \big\vert\alpha'^{vE}_{xy}\big\vert^{2} + \big\vert\alpha'^{vE}_{yy}\big\vert^{2} + \big\vert\alpha'^{EE}_{xy}\big\vert^{2} + \big\vert\alpha'^{HE}_{zy}\big\vert^{2} \le \frac{-8\text{Im}(1/\alpha'^{EE}_{yy})-1}{\big\vert1/\alpha'^{EE}_{yy}\big\vert^{2}} \le  16
\end{equation}

\begin{equation}
\label{maximum6}
\big\vert\alpha'^{pH}_{z}\big\vert^{2} + \big\vert\alpha'^{vH}_{xz}\big\vert^{2} + \big\vert\alpha'^{vH}_{yz}\big\vert^{2} + \big\vert\alpha'^{EH}_{xz}\big\vert^{2} + \big\vert\alpha'^{EH}_{yz}\big\vert^{2} \le \frac{-8\text{Im}(1/\alpha'^{HH}_{zz})-1}{\big\vert1/\alpha'^{HH}_{zz}\big\vert^{2}} \le  16,
\end{equation}
\end{subequations}
with the subscripts $x$, $y$, and $z$ denoting x-, y-, and z-component, respectively, from which it can be inferred that the maximum bounds are determined solely by the time frequency owing to the common product by the time frequencies in the normalized polarizability tensor of (\ref{polarizability_normalized}). Moreover, (\ref{maximum}) shows that the maximum bounds by the non-bianisotropic terms are reached at resonance with $1/\alpha'_{\text{diag}}=-i/4$ where $\alpha'_{\text{diag}}$ is the diagonal components in the normalized polarizability tensor of (\ref{polarizability_normalized}). These conditions generalize the effect of bianisotropic metamaterial properties on scattering, as the previous result in {\cite{quan2018maximum}} for the theoretical bounds of Willis acoustic scatterers is identically derived by making the electromagnetism-related terms zero and taking the time frequency out of the polarizability tensor. This indicates that the maximum scattering performance of bianisotropic metamaterials can be reached via the aid of EM coupling even for a simple spatial distribution that possesses nearly zero Willis coupling. It is noted that the derived conditions are general as the non-bianisotropic limit that is obtained with zero off-diagonal terms in the polarizability tensor can also be well recovered via the second RHS conditions of (\ref{maximum}) as
\begin{equation}
\text{Im}(1/\alpha'^{pp}),~ \text{Im}(1/\alpha'^{vv}_{xx}),~ \text{Im}(1/\alpha'^{vv}_{yy}),~ \text{Im}(1/\alpha'^{EE}_{xx}),~ \text{Im}(1/\alpha'^{EE}_{yy}), \text{ and } \text{Im}(1/\alpha'^{HH}_{zz}) \le -\frac{1}{8},
\end{equation}
indicating that the bianisotropic coupling terms can be of the same order as the non-bianisotropic terms in the scattering.

\vspace{0mm}
\begin{figure}[h]
\begin{center}
\includegraphics[width=0.8\columnwidth]{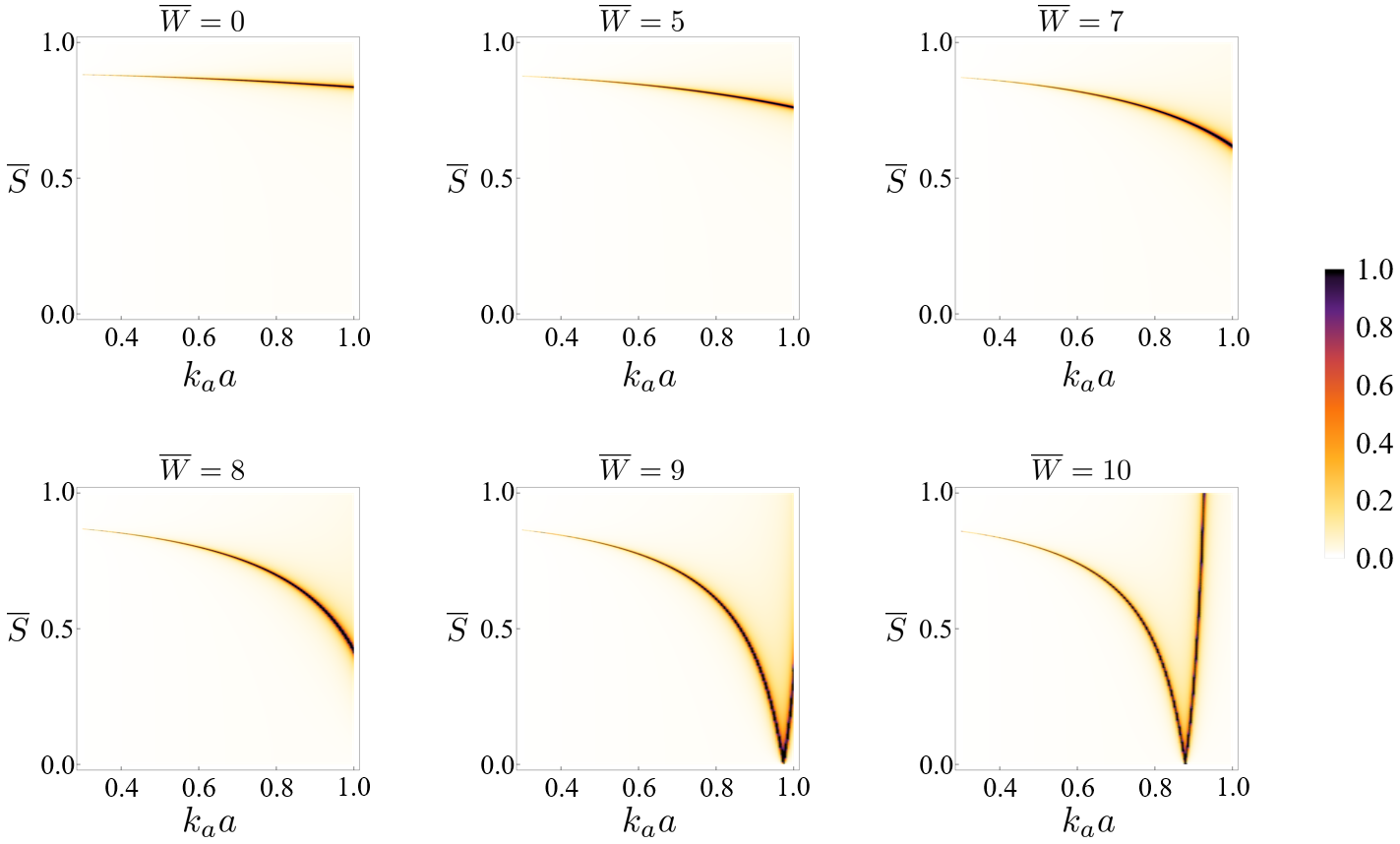}
\vspace{-2mm}
\caption{\label{fig1} Visualization of the bianisotropic scattering performance for different coupling values and scatterer sizes at $\theta_{s}=0$, using the first LHS of (\ref{maximum2}) normalized by 16. The colorbar denotes the normalized bianisotropic scattering performance.} 
\vspace{0mm}
\end{center} 
\end{figure}

For validation of these derived theoretical bounds, we analytically solve a scattering problem by considering a z-axis infinitely-long cylindrical particle with radius $a$ in the air, which possesses both Willis and EM coupling, under plane incident waves to be scattered. Here, acoustic and TE electromagnetic incident waves are simultaneously imposed on the particle to investigate the effect of the electromechanical bianisotropy of EM coupling on the scattering. Bianisotropic coupling effects associated with dynamics, such as Willis and EM coupling, result from two physical origins -- local coupling due to asymmetry of microstructures and nonlocal coupling due to heterogeneous material distribution {\cite{muhlestein2016reciprocity,norris2012analytical,nemat2011overall,alu2011first,sieck2015dynamic,torrent2015resonant,ponge2017dynamic}}. For lossless materials, the local and nonlocal contributions to the coupling effects in the constitutive equation of (\ref{constitutiveeq_macro}) are represented as purely real and imaginary parts, respectively. Even for lossy material systems, the nonlocal contributions may be neglected by properly designing a small, isolated metamaterial element rather than an ensemble of mutually interacting elements {\cite{muhlestein2017experimental}}. Similar to $\widetilde{S}^{jki}=\widetilde{S}^{\dagger ijk}$ {\cite{muhlestein2016reciprocity}}, the local reciprocity requires $\widetilde{W}^{ji}=\widetilde{W}^{\dagger ij}$. Here, we adopt the general coordinates with the standard notation that Latin indices $i,j,\ldots$ run from 1 to 3. Therefore, for this present scattering analysis, we assume that only the local contributions to the coupling effects are significant as $\widetilde{\bm{S}} = S\bm{d}_{s}\otimes\bm{d}_{s}\otimes\bm{d}_{s}$ and $\widetilde{\bm{W}}=W\bm{I}$ where $S$ and $W$ are the Willis and EM coupling coefficients, respectively, and $\bm{d}_{s}=\cos{\theta_{s}}\bm{e}_{x}+\sin{\theta_{s}}\bm{e}_{y}$ is the unit director vector with x- ($\bm{e}_{x}$) and y- ($\bm{e}_{y}$) direction unit vectors, and $\bm{I}$ is the second-order identity tensor. Fig.~\ref{fig1} shows representative results of the bianisotropic scattering performance using the first LHS of (\ref{maximum2}) for different coupling values and scatterer sizes, which verify the derived maximum theoretical bounds. Based on the directivity characteristics of the used Willis and EM coupling tensors, these coupling tensors are normalized as $\overline{S}=S/(i\sqrt{Y^{s}\rho^{s}})$ and $\overline{W}=W/(\sqrt{\epsilon_{em}^{s}\rho^{s}})$ where $Y^{s}$, $\rho^{s}$, and $\epsilon_{em}^{s}$ are Young's modulus, mass density, and electric permittivity of the scatterer, respectively. Furthermore, these results evidently show that Willis and EM couplings assist each other so that even with small Willis coupling values, the bianisotropic scattering performance can become strong and comparable with the direct non-bianisotropic terms. This further alludes to the added extra degree of design freedom to metamaterials that EM coupling can offer beyond just Willis coupling alone can achieve. 

\vspace{0mm}
\begin{figure}[h]
\begin{center}
\includegraphics[width=0.55\columnwidth]{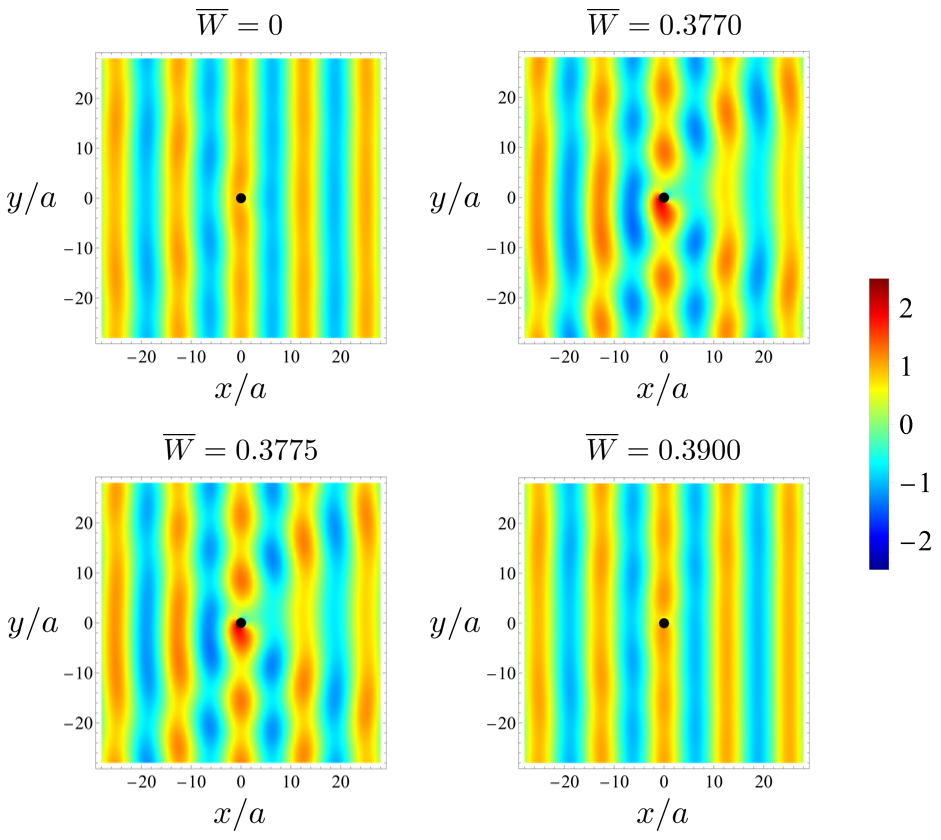}
\vspace{-2mm}
\caption{\label{fig2} Normalized acoustic pressure fields for the plane incident wave along the positive x-direction, for different EM coupling values at $\overline{S}$ = 0.8734 and $\theta_{s} = 0$ with $k_{a}a$ = 0.5. The black solid circles at the center denote the cylindrical scatterer. The colorbar denotes the acoustic pressure normalized by the incident pressure at the scatterer center, $p_{in_0}$.} 
\vspace{0mm}
\end{center} 
\end{figure}

Besides the scattering enhancement, we extend the idea, that EM coupling tailors the scattering phenomenon, into tunable wave scattering-cloaking devices in which the cloaking gives rise to object invisibility and/or immunity to detection. Fig.~\ref{fig2} shows the normalized acoustic pressure fields for the plane incident wave along the positive x-direction, for different EM coupling values at the specific Willis coupling and scatterer size. These different EM coupling values are chosen based on the maximum scattering performance. In the figure, the cylindrical scatterer is normalized by its radius $a$, denoted using black solid circles at the center, while the acoustic pressure field is normalized by the incident pressure at the scatterer center, $p_{in_0}$. Building on the limit applicable to the material homogenization scheme, i.e., characteristic size of metamaterials should be much less than wavelength, our target scatterers possessing coupling effects need to have a sufficiently small size for which scattering may barely occur such that the incident waves pass through without any perturbation when non-bianisotropic materials are considered. However, upon a certain value of Willis coupling, controlling EM coupling by modulation of external electric stimuli can give rise to comparable scattering phenomena with such small particles at the point reaching the derived maximum bounds. On the other hand, turning the external stimuli off or further increasing EM coupling removes the scattering again, i.e., tunable scattering-cloaking devices. We note that the shape of wave fields from the bianisotropic scattering is similar for different scatterer sizes when the scatterers have the same form of the coupling tensors, $\widetilde{\bm{S}}$ and $\widetilde{\bm{W}}$. This implies that the shape of scattered wave fields can also be engineered by designing scatterers to possess desired coupling tensors.

In summary, we derived tight, general theoretical bounds on the effect of bianisotropic coupling on piezoelectric metamaterial scatterers, based on the restrictions imposed by energy conservation. This analysis generalized two polarizabilities in elastodynamics and electromagnetics for piezoelectric materials based on their intrinsic electromechanical coupling characteristics. Our results show that for scattering performance, the bianisotropy induced by EM coupling can be of the same order as the non-bianisotropic terms in the polarizability tensor even for small-sized and simple metamaterials with small Willis coupling. Understanding these theoretical bounds can provide helpful information on the performance space of piezoelectric metamaterials for various applications such as acoustic sensing. This sheds light on the promising potential of piezoelectric metamaterials for tunable scattering devices whose bianisotropy can be modulated by external stimuli. It is envisioned that our study will support experimental and theoretical efforts to design optimal metamaterials with large EM coupling as well as to discover rigorous relations between EM coupling and electromagnetic waves which can be used as practical external stimuli and/or incident waves to be scattered, such as low-frequency waves of AC powers and radio-frequency waves of radars.

\begin{acknowledgments}
This research was supported by the Defense Advanced Research Projects Agency (Fund Number: W911NF2110363) and Office of Naval Research (Fund Number: N00014-21-1-2604). Additionally, the authors would like to acknowledge support from the Extreme Science and Engineering Discovery Environment (XSEDE) Bridges system, from National Science Foundation (Fund Number: ACI-1548562).
\end{acknowledgments}

\bibliography{biblio_scattering}

\end{document}